\documentstyle[12pt]{article}
\advance \topmargin by -\headheight
\advance \topmargin by -\headsep

\evensidemargin \oddsidemargin
\begin{document}

\topmargin -.6in

\def\rf#1{(\ref{eq:#1})}
\def\lab#1{\label{eq:#1}}
\def\nonu{\nonumber}
\def\br{\begin{eqnarray}}
\def\er{\end{eqnarray}}
\def\be{\begin{equation}}
\def\ee{\end{equation}}
\def\foot#1{\footnotemark\footnotetext{#1}}
\def\lb{\lbrack}
\def\rb{\rbrack}
\def\({\left(}
\def\){\right)}
\def\lskip{\vskip\baselineskip\vskip-\parskip\noindent}
\relax

\def\a{\alpha}
\def\b{\beta}
\def\d{\delta}
\def\D{\Delta}
\def\eps{\epsilon}
\def\g{\gamma}
\def\h{{1\over 2}}
\def\l{\lambda}
\def\L{\Lambda}
\def\m{\mu}
\def\n{\nu}
\def\o{\over}
\def\om{\omega}
\def\O{\Omega}
\def\pa{\partial}
\def\pr{\prime}
\def\ra{\rightarrow}
\def\s{\sigma}
\def\S{\Sigma}
\def\p{\phi}
\def\P{\Phi}
\def\jc{J^C}
\def\dj{{\cal D}}
\def\lie{{\cal G}}
\def\sw{w_{\infty}}
\def\bw{W_{\infty}}
\def\Tr{\mathop{\rm Tr}}

\def\rlx{\relax\leavevmode}
\def\inbar{\vrule height1.5ex width.4pt depth0pt}
\def\IR{\rlx\hbox{\rm I\kern-.18em R}}
\newcommand{\nit}{\noindent}
\newcommand{\ct}[1]{\cite{#1}}
\newcommand{\bi}[1]{\bibitem{#1}}
\begin{titlepage}

January, 1992 \hfill{UICHEP-TH/92-1}

\hfill{IFT-P/003/92-SAO-PAULO}

\vskip .6in

\begin{center}
{\large {\bf A New Deformation of W-Infinity
and Applications to the Two-loop WZNW and Conformal
Affine Toda Models}}
\end{center}

\normalsize
\vskip .4in

\begin{center}
{ H. Aratyn\footnotemark
\footnotetext{Work supported in part by U.S. Department of Energy,
contract DE-FG02-84ER40173 and by NSF, grant no. INT-9015799}}

\par \vskip .1in \noindent
Department of Physics \\
University of Illinois at Chicago\\
Box 4348, Chicago, Illinois 60680\\
\par \vskip .3in

\end{center}

\begin{center}
{L.A. Ferreira\footnotemark
\footnotetext{Work supported in part by CNPq}}, J.F. Gomes$^{\,2}$ and
A.H. Zimerman$^{\,2}$

\par \vskip .1in \noindent
Instituto de Fisica Te\'{o}rica-UNESP\\
Rua Pamplona 145\\
01405 S\~{a}o Paulo, Brazil
\par \vskip .3in

\end{center}

\begin{center}
{\large {\bf ABSTRACT}}\\
\end{center}
\par \vskip .3in \noindent

We construct a centerless W-infinity type of algebra in terms
of a generator of a centerless Virasoro algebra and an abelian
spin-1 current.
This algebra conventionally emerges in the study of
pseudo-differential operators on a circle or alternatively within KP
hierarchy with Watanabe's bracket.
Construction used here is based on a special deformation of the algebra
$w_{\infty}$ of area preserving diffeomorphisms of a 2-manifold.

We show that this deformation technique applies to the two-loop WZNW
and conformal affine Toda models, establishing henceforth $\bw$ invariance
of these models.

\end{titlepage}

\section{Introduction}

Extended (higher spin) symmetries of $\bw$-type are subject of a lot of
recent activities.
Bakas \ct{bakas} has called attention to the relevance of the area preserving
diffeomorphisms  on 2-manifold generating
$w_{\infty}$ algebra obtained by taking large $N$ limit of Zamolodchikov
$W_N$ algebra \ct{zamo}.
A deformation of $w_{\infty}$ allowing for central extensions
was later constructed by purely algebraic considerations in \ct{pope}.
According to the usual convention we refer to this deformation as
$\bw$ algebra.
There are two more independent constructions of the $\bw$ structure.
Namely, Moyal bracket acting on 2-dimensional functions \ct{moyal}
and the Lie algebra structure associated with the ring of
pseudo-differential operators on a circle \ct{radul}.
The general relations between these various realizations of
the $\bw$ symmetry has been analyzed in
\ct{fletcher,fairlie,igor,khesin}.

Recently we have found that the Conformal Affine Toda (CAT) model \ct{bb}
involves higher spin symmetry generators of arbitrary conformal spin
\ct{hspin}. These generators were constructed to be primary and as a
consequence they satisfied a nonlinear algebra.
In a special limit this nonlinear algebra reduces to $w_{\infty}$
algebra.
The most attractive feature of the CAT model is that it is conformal
invariant and there are reasons to expect it to be integrable.
It is therefore natural to ask whether interplay between conformal structure
and infinite many charges in involution will result in extended conformal
symmetry of $\bw$-type.
In this paper we indeed find a realization of $\bw$ algebra in terms
of infinitely many conserved charges within the simple Lagrangian structure
of the CAT model.

This paper presents two main results.
The first result described in section $2$ is a mathematical
construction based on a special deformation technique.
This construction starts with the area preserving diffeomorphism algebra
defined in terms of basic primary tensors $U$ and $J$ having respectively
spin 2 and 1.
In our setting $J$ is a self commuting current and $U$ is a generator of Witt
(centerless Virasoro) algebra.
By allowing a central element $h$ in the bracket between
$U$ and $J$ we are given a deformation parameter used to trigger our
deformation scheme.
As a result we obtain a realization of $\bw$ algebra in
terms of products of $U$ and differential polynomials of $J$, which is
isomorphic to the centerless algebra of pseudodifferential symbols
\ct{radul} with brackets between generators taking form \`{\it a} {\it la}
Watanabe \ct{watanabe}.

We emphasize that this construction can be realized without any particular
physical model in mind however it is suitable to describe (at least) part
of the symmetry structure of the CAT model.
In the section $3$ we show how our deformation scheme fits naturally into
the framework of two-loop WZNW and CAT models, which is the second main
result of this paper.
This result provides field theoretic realization, on the classical level,
of centerless $\bw$ in terms of interacting fields
\section{Deformation of the Algebra of the Area Preserving Diffeomorphisms}

The starting point of our construction are two fundamental objects given by
spin $2$ tensor $U$ and spin $1$ current $J$, obeying
the following basic Poisson brackets:
\br
\{ U (x) \, , \, U (y) \} &=& 2 U (y) \d^{\pr}
(x - y) -  U^{\pr} (y) \delta (x - y)
\label{eq:vira}\\
\{ U (x) \, , \, J (y) \} &=& J (y) \d^{\pr}
(x - y) -  J^{\pr} (y) \d (x - y) - h \d^{\pr\pr} (x - y)
\label{eq:wpi}\\
\{ J (x) \, , \, J (y) \} &=& 0 \lab{abel}
\er
with $h$ being a c-number and $\pa_y J (y) = J^{\pr} (y)$.

{}From the above Poisson brackets we derive easily
\be
\{ U ( x) \, , \, J^{n+1} (y) \}= ( n+1) J^{n+1} (y) \d^{\pr}
(x - y) -  \(J^{n+1} (y)\)^{\pr} \d (x - y) - h ( n+1) J^n (y) \d^{\pr\pr}
(x - y)
\lab{ujn}
\ee
Define now spin $n$ objects $V^n (x) \equiv U (x) J^{n-2} (x) $
with $n \geq 2$.
It is easy to see that for $h = 0$ the generators $V^n (x)$ satisfy the
classical $\sw$ algebra:
\be
\{ V^n (x) \, , \, V^m (y) \} = \( n + m -2 \) V^{n+m-2} (y) \d^{\pr} (x - y)
- ( n -1 ) \( V^{n+m-2 } (y)\)^{\pr} \d (x -y)
\lab{smallw}
\ee
Proof follows from \rf{vira} and \rf{ujn} by setting $h=0$.

In the following we ``deform" $\sw$ algebra in \rf{smallw}
by allowing $ h \ne 0$ in \rf{wpi} and \rf{ujn}.
For this purpose we introduce the differential operator:
\be
\dj \equiv h \pa + J
\lab{covariant}
\ee
with properties $\dj^{-1} J = 1$ and $\dj^{0} J = J$.
The differential operator $\dj$ satisfies $\dj (AB) = (\dj A) B + h A (\pa B)$,
which results in a modified Leibniz rule:
\be
\dj^n (AB) = \sum_{k=0}^{n} h^k {n \choose k} ( \dj^{n-k}A) ( \pa^k B)
\lab{leibniz}
\ee
leading to the following recurrence formula (setting $A=1$ and using
$\dj^n1=\dj^{n-1} J$):
\be
\dj^n = \sum_{k=0}^{n} h^k {n \choose k} \(\dj^{n-k-1}J (y)\) \pa^{k}
\lab{recurrence}
\ee
Another identity satisfied by differential polynomials of $\dj$ is:
\be
\sum_{k=0}^{m} (-h)^k {m \choose k} \pa^k \( \dj^{m+n-k} J \) = \sum_{k=0}^{m}
(-h)^k {m \choose k} \dj^nJ \pa^k \( \dj^{m-k-1}J\)
\lab{dd}
\ee
which follows from repeated application of the usual and modified
\rf{leibnitz} Leibniz rules.
We now introduce spin $n$ conformal objects
\be
W_n (x) = U (x) \dj^{n-3} (x) J (x) \qquad;\qquad n=2,3, \ldots
\lab{basic}
\ee
with lowest order generators given by:
\be
W_2 = U \;\; ;\;\; W_3 = V_3 =UJ \; \; ; \;\;
W_4 = U \( J^{2} + h J^{\pr} \) \; \; ;\;\;
W_5 =   U \( J^{3} + 3h J J^{\pr} + h^2 J^{\pr \pr} \) \;{\rm etc}.
\lab{ws}
\ee
One finds from \rf{recurrence} a generalized ``Miura transformation"
relating powers of $\dj$ to generators $W_k$:
\be
U \dj^n =\sum_{k=0}^{n} h^k {n \choose k} W_{n-k+ 2}\, \pa^{k}
\lab{miura}
\ee
One can prove that $W_n$ are quasi-primary generators satisfying
\br
\lefteqn{
\{ U (x) \, , \, W_n (y) \} = n W_n (y) \d^{\pr} (x -y)
- \( W_n (y) \)^{\pr} \d (x-y)} \nonu \\
&+& \sum_{k=1}^{n-2} (-1)^k h^k { n- 1 \choose k+1}
W_{n-k} (y) \, \d^{(1+k)} (x-y)
\lab{uwn}
\er
where $\d^{(k)} (x-y) = \pa^k \d (x-y) / \pa x^k$.

Furthermore using recurrence formula \rf{recurrence} we find an explicit
expression for:
\br
\lefteqn{
\{ W_3 (x) \, , \, W_n (y) \} = \( n + 1 \) W_{n+1} (y) \d^{\pr} (x - y)
- 2 W_{n+1}^{\pr} (y) \d (x -y) } \nonu\\
&+& { h \o 2!} \( 2 {\pa^2 \o \pa y^2} - (n-1) (n-2) {\pa^2 \o \pa x^2} \)
W_{n} (y) \d (x - y) \nonu \\
&+& \sum_{k \geq 3}^{n-1} (-h)^{k-1} {n-1 \choose k}
W_{n-k+2} (y) \d^{(k)} (x - y)
\lab{w3wn}
\er
which shows closure of this part of algebra.

The above result can be extended to the general Poisson bracket relation
for arbitrary generators $W_n$ and $W_m$ defined in \rf{basic}:
\br
\lefteqn{
\{ W_n (x) \, , \, W_m (y) \} = \( n +m -2\) W_{n+m-2} (y) \d^{\pr} (x - y)
- (n-1) W_{n+m-2}^{\pr} (y) \d (x -y) } \nonu\\
&+& \( \sum_{k =2}^{m-1} (-h)^{k-1} {m-1 \choose k} {\pa^k \o \pa x^k}
- \sum_{k =2}^{n-1} (-h)^{k-1} {n-1 \choose k} {\pa^k \o \pa y^k} \)
W_{n+m-k-1} (y) \d (x - y)
\lab{wnwm}
\er
by direct calculation using \rf{recurrence} and \rf{dd}.

As shown by Dorfman and Gelfand \ct{gelfand} (see also \ct{fletcher})
any 2-index infinite Lie algebra
may be written as a bracket algebra of functions.
In fact our algebraic relation \rf{wnwm} has a nice geometric realization
in terms of functions $e^{ipx} y^{n-1}$ on a cylinder $S^1 \times \IR$ with
a Lie algebra structure given by \ct{radul}
\be
\lb f (x,y) \, ,\, g (x,y) \rb = \sum_{k \geq 1} (-h )^{k-1} {1 \o k !} \(
{\pa^k f \o \pa x^k} {\pa^k g \o \pa y^k}
-{\pa^k f \o \pa y^k} {\pa^k g \o \pa x^k} \)
\lab{symbol}
\ee
which is a commutator $\lb f , g \rb = f \circ g - g \circ f$
with respect to the product of symbols:
\be
f (x,y) \circ g (x,y) = \sum_{k \geq 0} (-h )^{k-1} {1 \o k !}
{\pa^k f \o \pa x^k} {\pa^k g \o \pa y^k}
\lab{product}
\ee
The connection becomes transparent by noticing that the mapping
$W_n^p \leftrightarrow e^{ipx} y^{n-1}$ where
$W_n (x) = \sum_{p} W_n^p e^{ipx}$ is an isomorphism between \rf{wnwm}
and the algebra of $e^{ipx} y^{n-1}$  under the
bracket \rf{symbol}.
This connection establishes among other
things that \rf{wnwm} satisfies Jacobi relations.

Less technical argument for validity of \rf{wnwm} for the
generators given by formula \rf{basic}
is based on the following Jacobi identity (which is automatically
satisfied in our case)
\be
\{ W_n (x)  \, , \,\{ W_3 (y)  \, , \, W_3 (z) \} \} =
- \{ W_3 (z) \, , \, \{ W_n (x) \, , \, W_3 (y)\} \}
-\{ W_3 (y) \, , \, \{ W_3 (z) \, , \, W_n (x)  \} \}
\lab{jac}
\ee
The terms on the right hand side of \rf{jac} contain only brackets of $W_3$
and $W_m$ which are always given by \rf{w3wn} as found above.
The term on the left hand side involves as the ``highest" bracket
$\{ W_n (x) \, , \, W_4 (y) \} $, which therefore can uniquely be written as
\underbar{linear} combination of $W_n$'s.
Since this is a well-defined and \underbar{unique} solution to the algebraic
identity \rf{jac} it must agree with \rf{wnwm}.
We can now recursively extend this argument to
$\{ W_n (x) \, , \, W_5 (y) \} $ etc. etc.
Therefore, in general, knowing the bracket $\{ W_n (x) \, , \, W_m (y) \} $
for $m <n$ using Jacobi identity for
$\{ W_n (x)  \, , \,\{ W_3 (y)  \, , \, W_m (z) \} \}$ we obtain the unique
expression for $\{ W_n (x) \, , \, W_{m+1} (y) \} $ in terms of lower order
brackets.
This concludes our induction argument for \rf{wnwm}.

We also observe that algebra \rf{wnwm} is of the Watanabe \ct{watanabe} form.
To see it we make the identification $W_{n+1} \to u_n $, which results in:
\br
\{ u_n (x) \, , \, u_m (y) \} &=& \O_{nm} (x) \, \d (x-y) \qquad\qquad
\qquad n,m \geq 1 \lab{wata} \\
\O_{nm} (x) &=& \sum_{k=0}^{n} h^{k-1} { n \choose k} u_{n+m-k} (x) \pa_x^k
-\sum_{k=0}^{m} (-1)^k  h^{k-1} { m\choose k} \pa_x^k u_{n+m-k} (x)
\nonu
\er
This identification is not surprising in view of the fact that \rf{wnwm}
is isomorphic to the commutator \rf{symbol} for pseudodifferential operators
on a circle, see also \ct{wu}.

\section{W-Infinity Symmetry of the two-loop WZNW and Conformal Affine Toda
Models}

As recently shown \ct{toda} the conformal affine Toda (CAT) models can be
obtained via Hamiltonian reduction from a two-loop Kac-Moody (KM) algebra.
This reduction preserves conformal invariance, here we will show that it also
maintains W-infinity symmetry.
The structure of the representations of the two-loop KM algebra
has been further analyzed in \ct{schwimmer}.

In reference \ct{toda} we have constructed the classical Sugawara
energy-momentum tensor:
\be
T (x) = {1 \o 2 } \Bigl( \sum_{a,b=1}^{{\rm dim}\, \lie}
\sum_{n=- \infty}^{\infty} g^{ab} J^n_{ a} (x) J^{-n}_{b} (x)  +
2 J^{D}(x) J^{C} (x) \Bigr) \label{eq:suga}
\ee
in terms of the currents $J^n_{ a} (x)$, $J^{D}(x)$ and $J^{C} (x)$
satisfying two-loop KM current algebra (equations (2-5) in \ct{toda}).

First let us note that $T (x)$ and $J^{C} (x)$ satisfy algebra given in
\rf{vira}-\rf{abel} with $h=0$ generating therefore $\sw$ symmetry of two-loop
WZNW model.
In fact this model has a bigger symmetry as we will show now
by first defining the modified energy-momentum tensor:
\be
L (x)
= T (x) + \pa_x \( 2 J_{\hat \delta} (x) +
h J^{D }(x) \)
\label{eq:modsuga}
\ee
where $J_{\hat \delta}=k \Tr \( {\hat g}^{-1} \pa_{+} {\hat g} \,{\hat
\d}\cdot H^{0} \)$ with ${\hat \d} = \h \sum_{\a>0} \a /\a^2$ and $h$
is the Coxeter number of the underlying semisimple Lie algebra
(in case of $sl (2)$ $h=2$). For missing notational explanations and more
details see \ct{toda}.

The modified energy-momentum tensor \rf{modsuga} was introduced in \ct{toda}
in order to ensure
conformal invariance of our Hamiltonian reduction scheme.
We observed there that $L (x)$ satisfies the Virasoro algebra:
\be
\{ L (x) \, , \, L (y) \} = 2 L(y) \d^{\pr}
(x - y) -  L^{\pr} (y) \delta (x - y) - c \d^{\pr\pr\pr} (x - y)
\label{eq:w2w2}
\ee
with the central element $c = 4 {\hat \d}^2$.
Furthermore we find:
\be
\{ L (x) \, , \, J^C (y) \}= J^C (y) \d^{\pr}
(x - y) - \( J^C (y)\)^{\pr}  \d (x - y) - h \d^{\pr\pr} (x - y)
\label{eq:w2jc}
\ee
It is important at this point to notice that the central element in \rf{w2w2}
can be removed by redefining the energy-momentum tensor according to:
\be
U (x) \equiv L(x) + \g \pa_x \jc (x)
\lab{utensor}
\ee
It is easy to verify that $U (x)$ satisfies \rf{w2w2} with c-term
equal to $\( 2h \g - c \) \d^{\pr\pr\pr} (x - y) $.
Anomaly free Virasoro algebra is obtained by choosing $\g = c / 2 h$.
Bracket between the energy-momentum tensor and $\jc (x)$ \rf{w2jc}
is unchanged by this last modification.

Now we turn our attention to the CAT model.
The Lagrangian density for this model with the finite semisimple Lie algebra
$\lie$ is given by
\be
{\cal L} = \h \sum_{a,b=1}^{{\rm rank}\, \lie}
\eta_{ab} \pa_{+} \phi^a \pa_{-} \phi^b + \pa_{+} \mu \pa_{-} \nu
+ \pa_{-} \mu \pa_{+} \nu + 2 \sum_{a=1}^{{\rm rank}\, \lie}
{ 1 \o \a^2_a} e^{K_{ab} \phi^b} + {2 \o \psi^2} e^{- K_{\psi b}
\phi^b + 2 \m}
\label{eq:lagrang}
\ee
written in the light-cone coordinates $x_{\pm} = x \pm t$, $\pa_{\pm} =
\h ( \pa_{x} \pm \pa_{t})$.

The above Lagrangian density reproduces equations of motion
obtained via Hamiltonian reduction from 2-loop WZNW model
\br
\pa_{-} \pa_{+} \phi_a & =& \,e^{K_{ab} \phi_b}
- \,e^{- K_{\psi b} \phi_b + 2 \mu}
\label{eq:todaone} \\
\pa_{-} \pa_{+} \nu &=&  {2 \o \psi^2}
\, e^{- K_{\psi b} \phi_{b} + 2\mu} \quad;\quad
\pa_{-} \pa_{+} \mu = 0  \label{eq:todathree}
\er
In equations \rf{lagrang}-\rf{todathree}
we introduced $\eta^{ab} = {\a^2_b \o 2} (K^{-1})^{ab} $
with $K_{ab} = 2 \a_a \cdot \a_b /\a_b^2$ being the Cartan matrix for
$\lie$, $\a_a$ denote the simple roots, while $\a, \b$ denote arbitrary
roots of $\lie$ and $\psi$ is the highest root of $\lie$
(see \ct{toda} for more details).

The Lagrangian density \rf{lagrang} for the case
of $\lie = sl (2)$ was originally introduced in \ct{bb} and writes as
(normalizing $\alpha^2 = 2$):
\be
{\cal L}_{sl (2)} = \pa_{+} \phi \pa_{-} \phi + \pa_{+} \mu \pa_{-} \nu
+ \pa_{-} \mu \pa_{+} \nu + e^{2 \phi} + e^{- 2 \phi + 2 \m}
\label{eq:sl2lagr}
\ee

All the subsequent calculations are applicable for both chiralities,
here we only write results for the right chirality.
{}From \rf{lagrang} we find the canonical momenta with respect to ``time"
$x_{-}$:
\be
\pi_a = \h \eta_{ab} \pa_{+} \p^b \quad;\quad \pi_{\m}= \pa_{+} \n
\quad;\quad \pi_{\n} = \h J^C = \pa_{+} \m
\lab{momenta}
\ee
where the subscripts $\m,\n$ are not the Lorentz
indices but refer to the corresponding fields.
{}From now on $x$ and $y$ stand for $x_{+}$ and $y_{+}$ and we drop the
subscript ``$+$" from all derivatives.

The usual canonical Poisson structure gives rise to the following Dirac
brackets for the constrained momenta \rf{momenta} for equal $x_{-}$:
\be
\{ \pi_a (x) \, , \, \pi_b (y) \} = {1 \o 4} \eta_{ab} \d^{\pr} (x -y)
\quad;\quad
\{ \pi_{\m} (x) \, , \, J^C (y) \} = \d^{\pr} (x -y)
\lab{dibra}
\ee
The model defined by \rf{lagrang}
is conformally invariant with the conserved and traceless generator
of conformal transformations given by
\be
L_{CAT} =  2 \sum_{a,b=1}^{{\rm rank}\, \lie}
\eta^{ab} \pi_a \pi_b + \pi_{\mu} J^C - 2
\sum_{a,b=1}^{{\rm rank}\, \lie} { 2 \o \a^2_a} \eta^{ab} \pa \pi_b
-  h \, \pa \pi_{\m}
\label{eq:wtwo}
\ee
Using the bracket relations \rf{dibra} we derive the corresponding
Virasoro algebra \rf{w2w2}.
Furthermore we derive from \rf{dibra} a relation \rf{w2jc}.
We now identify a conserved spin-2 tensor $U (x)$,
which satisfies an anomaly free Virasoro algebra within the CAT model
by the same procedure as in \rf{utensor}.

We therefore see that construction of section $2$ applies straightforwardly
to the two-loop WZNW and CAT models with the Coxeter number $h$ playing
a role of deformation parameter and $U$, $\jc$ coinciding with $W_2$
and $J$ of section $2$ establishing $W_{\infty}$ invariance of these models.
\lskip
{\em In Conclusion:} \\
We revealed the $\bw$ symmetry structure of the two-loop WZNW and CAT model
making use of deformation of algebra of the symplectic diffeomorphisms.
The algebra of conserved generators in \rf{wnwm} was shown to be isomorphic to
the Lie algebra structure of pseudodifferential symbols associated with a
cylinder.
This symmetry structure provides an explicit method of constructing
infinitely many conserved, local charges in involution in the usual
Lax formalism within KP hierarchy \ct{watanabe,wu,lax}.

Generalizations of the above deformation technique to include c-anomaly
and connection to non-linear realization of the algebra in terms of
primary field generators \ct{hspin} are under investigation.
\lskip
{\bf Acknowledgements:}\\
One of us (AHZ) acknowledges CNPq for financial support within CNPq/NSF
Cooperative Science Program and thanks Physics Department at the University of
Illinois at Chicago for hospitality.

\end{document}